\documentclass{emulateapj}
\slugcomment{Accepted for publication to ApJ}
\def\bmath#1{\mbox{\boldmath $#1$}}
\shorttitle{Magnetic Field Decay in the Outer Crust of Neutron Stars}
\begin{document}
\title{Magnetic Field Decay Due to the Wave-Particle Resonances in the
Outer Crust of the Neutron Star}

\author{Hiroyuki R. Takahashi\altaffilmark{1}, 
Kei Kotake\altaffilmark{1,2}, and Nobutoshi Yasutake\altaffilmark{2}}
\email{takahashi@cfca.jp}
\altaffiltext{1}{Center for Computational Astrophysics, National
  Astronomical Observatory of Japan, Mitaka, Tokyo 181-8588, Japan}
\altaffiltext{2}{Division of Theoretical Astronomy, National
  Astronomical Observatory of Japan, Mitaka, Tokyo 181-8588, Japan}
\begin{abstract}
 Bearing in mind the application to the outer crust of
the neutron stars (NSs), we investigate the magnetic field decay 
by means of the fully relativistic Particle-In-Cell simulations. 
 Numerical computations are carried out in 2-dimensions, 
in which the initial magnetic fields are
 set to be composed both of the uniform magnetic fields that model the global 
 fields penetrating the NS and of 
the turbulent magnetic fields that would be originated from the Hall cascade 
of the large-scale turbulence.
 Our results show that the whistler cascade of the turbulence transports 
the magnetic energy preferentially in the direction perpendicular to the uniform
magnetic fields.
 It is also found that the distribution function of electrons 
becomes anisotropic because electrons with lower energies are 
predominantly heated in the direction parallel to the uniform magnetic fields due to 
the Landau resonance, while electrons with higher energies are heated mainly 
by the cyclotron resonance that makes the distribution function isotropic
for the high energy tails. 
 Furthermore we point out that the degree of anisotropy takes maximum 
as a function of the initial turbulent magnetic energy.
As an alternative to the conventional ohmic dissipation, we propose that 
  the magnetic fields in the outer crust of NSs, 
cascading down to the electron inertial scale 
via the whistler turbulence, would decay predominantly by the 
dissipation processes through the Landau damping and the cyclotron resonance. 
\end{abstract}
\section{Introduction}\label{intro}
Pushed by the accumulating observations of radio pulsars, 
accreting neutron stars (NSs), and most recently, magnetars, extensive studies have been 
carried out to understand the evolution of magnetic fields in neutron
stars \citep[e.g.,][]{batta96,harding06,reisen09}.
The known radio pulsars so far are generally 
categorized into two classes: young ($\lesssim 10^7$ yr) pulsars with
spin periods $P\simeq 10^{-1\sim1}~\mathrm{s}$ and the magnetic field
strength $B\simeq 10^{10\sim 13}~\mathrm{G}$, and 
 the old millisecond radio pulsars, which have magnetic field strength as low as
 $10^{8\sim 9}~\mathrm{G}$. While most radio pulsars are isolated
 objects, the millisecond pulsars are 
predominantly in binaries, suggesting that the magnetic fields decay with time, 
perhaps by an accretion of matter from the binary companion 
\citep[e.g.,][and references therein]{harding06}.
 \begin{figure*}
 \begin{center}
 \includegraphics[width=13cm]{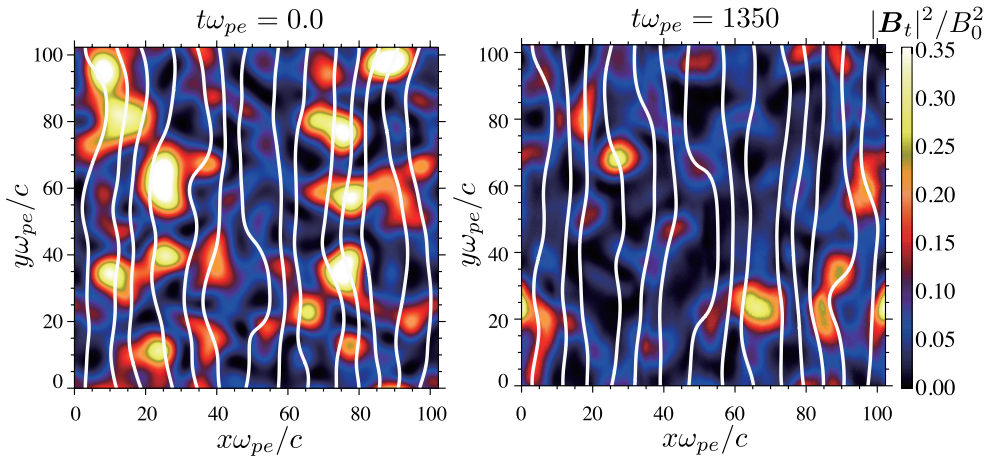}
  \caption{Turbulent magnetic field energy density $|\bmath
  B_t/B_0|^2= |(\bmath B - B_0 \bmath e_y)/B_0|^2$ (color) and the
  magnetic field lines (curves) at the initial state (left) and
  at $t\omega_{pe}=1350$ (right) for $\epsilon=0.1$.  }
  \label{fig:fig1}
 \end{center}
\end{figure*}

For the isolated radio pulsars, it is still an open question whether the
 NS magnetic fields do or do not decay with time. For example, 
\citet{narayan90} argued 
 that the field should decay exponentially on a few Myr timescale 
\cite[see also][]{cordes98,gont04}, while similar studies implied the decay time 
$\gtrsim 100$ Myr \citep[e.g.,][]{batta92,fauch06}. Such divergent results may come from the
uncertainties inherent to those population 
synthesis studies, such as the selection effects, the luminosity
evolution,
and the dependence of beaming fraction on period. 
 On the other hand, the discovery of magnetars has provided 
several important evidences to favor the magnetic field decay on relatively 
short timescales \citep[e.g.,][]{arras,harding06}.
In fact, the soft gamma-ray repeaters and anomalous X-ray pulsars, 
observed as young ($\lesssim 10^{4}$ yr) NSs with strong
($\sim 10^{14 \sim 10}~\mathrm{G}$) magnetic fields, are considered to
be powered by the decay of their magnetic fields \citep[e.g.,][]{woods}.
Giant flaring activities are proposed to be the rapid release of the 
 magnetic stress building in the NS crust
 \citep{thompson95, thompson96, 2006MNRAS.367.1594L}.
 More recently, \citet{pons} 
 presented an evidence that the magnetic field decay of $\sim 10$ Myr, can 
 explain the thermal evolution from magnetars continuously 
to ordinary radio pulsars. At present, these ideas and new observations seem to favor
 the existence of the magnetic field decay in the isolated NSs.

The pioneering study by \cite{1992ApJ...395..250G} identified the dissipation processes 
of the magnetic energy in the crust of the isolated NS during its evolution.
 They first showed that magnetized turbulence in the crust of NSs can be
 described by the electron magnetohydrodynamic (EMHD) equation, in which the 
 time evolution of the magnetic field is governed by the 
 advection of the field by the Hall drift, the Ohmic dissipation, and
 the ambipolar diffusion.
 On top of the Ohmic decay and the ambipolar diffusion, they first proposed that 
the Hall drift, though non-dissipative itself, could be an important ingredient for
 the field decay because it can lead to 
dissipation through the whistler cascade of the turbulence. 
As the eddy size of the turbulence becomes smaller, the magnetic energy
was transported from large to small scales, leading to the dissipation
of the magnetic energy finally via the Ohmic dissipation. 

In order to confirm their prediction, numerical simulations in EMHD are
required because such a cascading process is essentially a non-linear process. 
In the two- and three-dimensional (2D and 3D) simulations of the low
$\beta$ plasma, \citet{biskamp} showed a clear cascade of the energy 
due to the whistler turbulence over more than an order of magnitude in the
length scale. 
The 3D EMHD simulations by \cite{Cho04,Cho09} 
confirmed the scale-dependent anisotropy in the EMHD turbulence.
 They also showed that the anisotropic cascading processes in
 the EMHD proceed via the propagation of the whistler waves 
along the global magnetic fields \citep[see, also][]{2010AnGeo..28..597N}.
 This nature is in contrast to the conventional MHD turbulence, 
in which the Alfv\'en waves play a major role 
 to transport the turbulent magnetic energy 
\citep[e.g.,][]{1995ApJ...438..763G, 1989PhFlB...1.1964B,
2002ApJ...566L..49C}.
 Reflecting the fact that the propagation of the whistler waves are more dispersive 
than the Alfv\'en waves,
 the power spectral density in the EMHD turbulence was 
 shown to become steeper compared to the MHD turbulence. 
 Here it should be noted in the EMHD simulations that one can precisely 
follow the cascading process, however the dissipation process should be treated 
phenomenologically, i.e., via the resistivity or the so-called hyperdiffusivity
\citep{Cho04,Cho09}.

In this paper, we perform the Particle-In-Cell (PIC) simulations aiming to 
understand the decay process of the magnetic fields in NSs.
 As well known, the PIC simulations, which have been often
 performed to study the turbulent cascade in solar 
wind \citep{2008PhPl...15j2305S,2008GeoRL..3502104G,2010ApJ...716.1332G},
  can precisely take into account the 
electromagnetic modes in plasma by solving the full Maxwell equations coupling with 
the all species of plasma particles. 
The ordinary PIC simulation can treat the collisionless plasma in which 
the collisional frequency is smaller than the typical frequencies of 
plasma particles, such as the plasma frequency and the 
gyro frequency. \cite{2006PhRvD..74d3004S} have estimated the
collisional frequency between the particles in NSs. From
their results, the electron collisional frequency in the outer crust of
the NS is of the order of 
$10^{-2}\omega_{pe}$ with $\omega_{pe}$ being the electron plasma frequency, 
which indicates that the electrons are marginally
collisionless in the electron inertial scale. Since the gyro frequency 
is of the order of the plasma frequency, we approximately treat the plasma as
collisionless to study the decay process of the magnetic fields 
in the outer crust of the NS. As a complement to the foregoing 
EMHD simulations 
\citep[e.g.,][]{biskamp,Cho04,Cho09} that focus on the 
whistler cascade of the turbulence,
the PIC simulations that we will present in this paper, should merit 
for understanding the mechanism of the magnetic field decay,
cascading further down to the electron inertial scale.

This paper is organized as follows. 
In Section 2,  we show the numerical setup 
of the PIC simulations.  The numerical results of the turbulent cascade 
and the dissipation process are shown in \S~3.
 We summarize our results and discuss their implications in Section 4.

\section{Numerical Methods and Initial Conditions}\label{setup}
\begin{figure}
 \begin{center}
 \includegraphics[width=8cm]{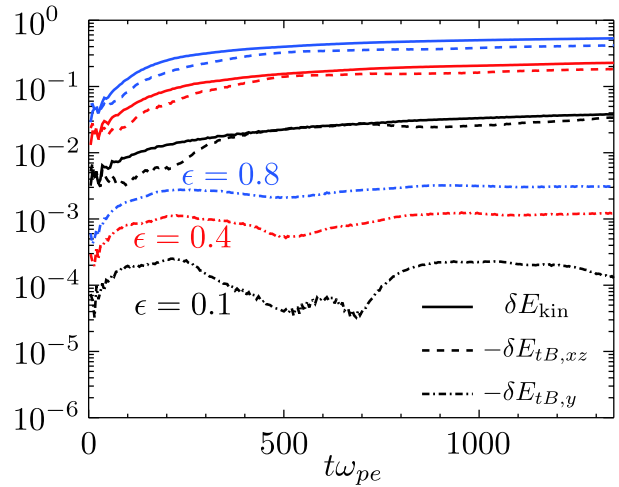}
  \caption{Time evolution of the energy deviation. 
  Blue, red, black curves are respectively for $\epsilon = 0.8, 0.4,
  0.1$ and the solid, dashed, dot-dashed curves denote the electron
  kinetic energy ($\delta E_\mathrm{kin}$), the minus of the turbulent magnetic
  field energy whose field component is perpendicular to $\bmath B_0$
  ($-\delta E_{tB,xz}$), and the minus of the turbulent magnetic field
  energy whose field
  component is parallel to $\bmath B_0$ ($-\delta E_{tB,y}$),
  respectively. }
  \label{fig:fig2}
 \end{center}
\end{figure}
We carry out 2-dimensional fully relativistic PIC
simulations \citep[e.g.,][]{2001plph.book.....L}. The basic equations
are described as,
\begin{figure*}
 \begin{center}
 \includegraphics[width=16cm]{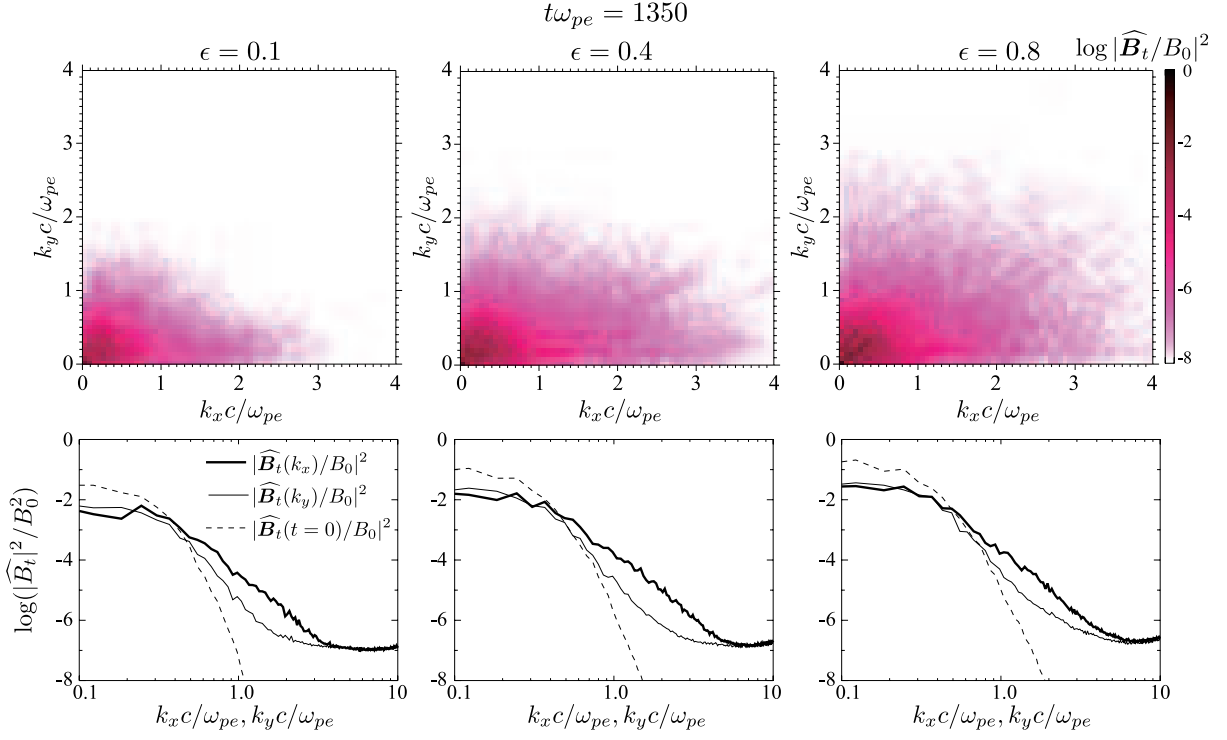}
  \caption{(Upper panel) Power spectral density of the turbulent
  magnetic fields, $\log |\widehat{\bmath B_t}/B_0|^2$ at
  $t\omega_{pe}=1350$
  for $\epsilon = 0.1, 0.4, 0.8$ from left to right, respectively.
  The lower panels show the 1-dimensional, cumulative power spectral
  density. Thick and thin solid curves show $|\widehat{\bmath B_t}(k_x)/B_0|^2$ and
  $|\widehat{\bmath B_t}(k_y)/B_0|^2$, respectively, while dashed curves do the initial
  spectra of the turbulent magnetic energy. }
  \label{fig:fig3}
 \end{center}
\end{figure*}
\begin{equation}
 \frac{\mathrm{d} \bmath p_i}{\mathrm{d}t} 
= q_i \left(\bmath E + \mbox{\boldmath$\beta$}_i \times \bmath B\right)
\end{equation}
\begin{equation}
 \nabla \cdot \bmath{B} = 0,
  \label{eq:divB}
\end{equation}
\begin{equation}
 \nabla \cdot \bmath E = 4\pi \rho_e, 
  \label{eq:divE}
\end{equation}
\begin{equation}
 \frac{\partial \bmath B}{\partial t} +  c \nabla \times \bmath E = 0,
  \label{eq:faraday}
\end{equation}
\begin{equation}
 \frac{\partial \bmath E}{\partial t} -  c \nabla \times \bmath B = -4\pi
  \bmath j,
  \label{eq:ampere}
\end{equation}
\begin{equation}
 \rho_e =  \sum_i q_i S(\bmath x-\bmath x_i),\label{eq:charge}
\end{equation}
\begin{equation}
 \bmath j = \sum_i q_i \bmath v_i S(\bmath x-\bmath x_i),\label{eq:current}
\end{equation}
where $\bmath p_i$, $q_i$, $\bmath E$, $\bmath B$, $\rho_e$, and
$\bmath j$  are the momentum, the particle charge,
the electric fields, the magnetic fields, the charge density, and the
current density, respectively. $\bmath \beta_i \equiv
\bmath v_i/c$ is the three velocity normalized by the speed of light. The
subscript $i$
denotes the particle species.
Here we ignore the quantum effects such as the Landau level, for simplicity. 
The particle motion is determined by solving the special relativistic
equation of motion. Then the electric charge and the charge current are 
obtained from equations (\ref{eq:charge}) and (\ref{eq:current}), in which 
 the shape factor $S$ can extrapolate the physical quantities on
 discretized grids at
$\bmath x$ from particles.
By using the electric charge
and the current density, the
electromagnetic fields are updated by solving Maxwell equations, so
that the system can be solved self-consistently.
We solve these equations in the rectangular coordinates by assuming
$\partial/(\partial z)=0$ (the so-called 2.5-dimension). 
Number of grid points is $(N_x, N_y)=(1024, 1024)$ and the
corresponding system size is $L_x = L_y = 102.4c/\omega_{pe}$,
where $\omega_{pe}\equiv\sqrt{4\pi n e^2/m_e}$ is the electron plasma frequency.
Each cell contains $80$ particles and the total
number of particles in the simulation box is $8.4\times 10^{7}$ for each species. 
The boundary conditions are periodic in both directions.
We assume that ions are immovable, which is reasonable in the outer
crust of the NS \cite[e.g.,][]{1992ApJ...395..250G}. 
We confirmed that the numerical results are
qualitatively and quantitatively consistent with those when we take into
account the ion motion.

In the initial state, the plasma distributes uniformly in space and 
the electron distribution function obeys the relativistic Maxwellian with the
temperature $T_e = 10^{8}~\mathrm{K}$ \cite[e.g.,][]{agui}. 
We assume that there are two components 
 of the magnetic fields in the NS. The first one is the uniform
 magnetic fields that would globally penetrate the NS, which should be 
 necessary to explain the spin-down of the NS. The global field
 is set to be uniform in space $\bmath B = B_0 \bmath e_y$, where
 $\bmath e_y$ is the unit vector in $y$-direction. 
 The corresponding ratio of the gyro frequency evaluated from the uniform magnetic fields $\omega_{ce}=eB_0/(m_ec)$ and
the plasma frequency $\omega_{pe}$ is $0.5$.
The second one is the turbulent magnetic fields, which would be cascaded
down from large to small scales due to the Hall turbulence. The initial
turbulent magnetic fields are modeled to take the form of 
$\widehat{\bmath B}(\bmath k)\propto \exp[-\bmath k^2/k_0^2+i\alpha_k]$,
where $\bmath k$ is the wave number,  $k_0 =0.31\omega_{pe}/c$ is
the typical wave number of the initial fluctuations and $\alpha_k$
is the random phase. The hat denotes the variable in
the Fourier space. We change the amplitude of the
turbulent magnetic fields $\bmath B_t \equiv \bmath  B - B_0\bmath e_y$
by introducing a parameter, $\epsilon \equiv
\int \mathrm{d}V~ \bmath B_t(t=0)^2/\int \mathrm{ d}V~B_0^2 = 0.1, 0.4, 0.8$. We also carry
out the simulation with $\epsilon = 0.0$ to assess the validity of our
simulation codes.

\section{Results}
\subsection{Energy Exchange}
Figure~\ref{fig:fig1} shows the turbulent magnetic field energy density (color) and
the magnetic field lines (white curves) at the initial state
$t\omega_{pe}=0$ (left) and at the end of the simulation
$t\omega_{pe}=1350$ (right) for $\epsilon = 0.1$. It is clearly shown that
the turbulent magnetic field energy given at the initial state (bright spots
 in the left panel) decreases at the end of the simulations (right panel). 
The pattern size of the turbulent
magnetic fields becomes relatively smaller with time. These
results imply that the magnetic energy is converted to the plasma
energy. 
\begin{figure}
 \begin{center}
 \includegraphics[width=7cm]{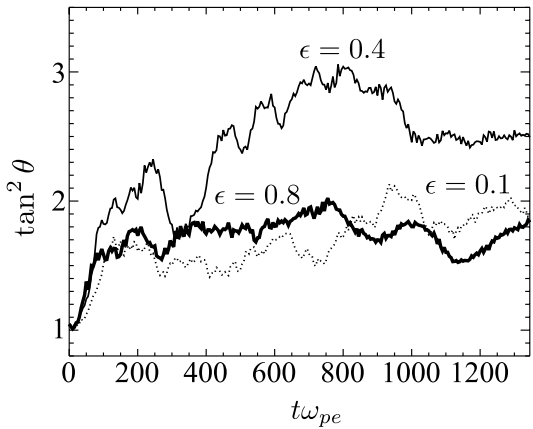}
  \caption{Time evolution of the anisotropy $\tan^2\theta$ for $\epsilon
  = 0.8$ (Thick solid), $\epsilon = 0.4$ (thin solid), and $\epsilon =
  0.1$ (dotted curve). } 
  \label{fig:fig4}
 \end{center}
\end{figure}

To see clearly how the energy conversion proceeds, we calculate
deviations of each energy from the initial state, such as the electron kinetic
energy $\delta
E_\mathrm{kin}\equiv\sum_i m[\gamma(t)-\gamma(t=0)]/E_{0}$ and 
 the turbulent magnetic field energies $\delta E_{tB,xz}=\int
 \mathrm{d}V~\{(B_x^2+B_z^2)-[B^2_x(t=0)+B^2_z(t=0)]\}/(8\pi E_{0})$ and $\delta
 E_{tB,y}=\int \mathrm{d}V~[B_y^2-B^2_y(t=0)]/(8\pi E_{0})$, 
where $E_0=\int dV B_0^2/(8\pi)$ is the magnetic energy of uniform fields.
 Figure \ref{fig:fig2} depicts the time evolution of the energy
 deviation of the electrons $\delta E_\mathrm{kin}$ (solid curves) and of the
 minus of the turbulent magnetic energies $-\delta E_{tB,xz}$ (dashed curves) and
 $-\delta E_{tB,y}$ (dot-dashed curves).  
 It can be seen that the turbulent magnetic energy in the perpendicular direction 
($\delta E_{tB, xz}$) decreases with time (note the minus sign), 
while the particle kinetic energy ($\delta E_\mathrm{kin}$) increases with it.
 The energy deviation of the particles $\delta E_\mathrm{kin}$ is almost
comparable to $-\delta E_{tB,xz}$. This means that the perpendicular component of the
magnetic energy is preferentially converted to the particle energy. 
 Here it should be noted that the small difference between $\delta
 E_\mathrm{kin}$ and $-\delta E_{B,xz}$ comes from the electric field
 energy generated by turbulent motions. The particle energy distribution
 function obtained 
in this simulation can be well fitted by the Maxwellian
distribution. Therefore,
the plasma is not accelerated, but is heated up through the particle-wave interaction.
 The resulting heating is expected to become larger 
 for models with larger initial turbulent magnetic energy (see models with different $\epsilon$ in Figure 2). The efficient energy conversion to the perpendicular component
 also suggests that the energy dissipation will proceed anisotropically, which 
 acts to make the configuration of the magnetic-field lines approach to the uniform
 one, i.e., the potential field. 

\subsection{Anisotropy}{\label{sub:anisotropy}}
\begin{figure}
 \begin{center}
 \includegraphics[width=7cm]{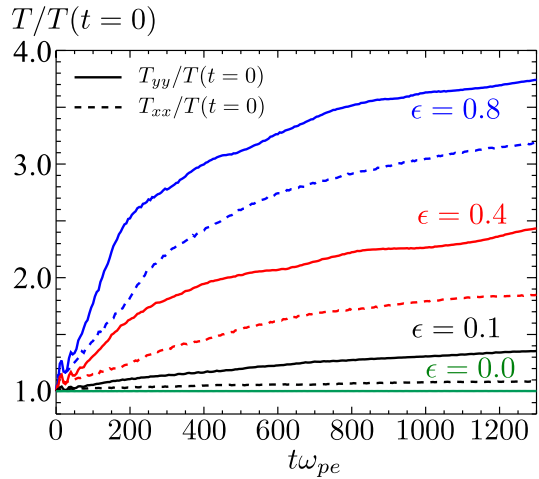}
  \caption{Time evolution of the electron temperature. 
  Blue, red, black, and green curves are, respectively, for $\epsilon =
  0.8, 0.4, 0.1, 0.0$ and solid and dashed curves are for $T_{yy}$ and
  $T_{xx}$, respectively. 
  } 
  \label{fig:fig5}
 \end{center}
\end{figure}

\begin{figure}
 \begin{center}
 \includegraphics[width=7cm]{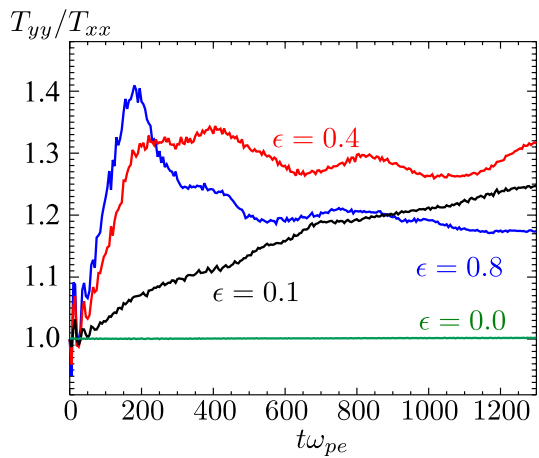}
  \caption{Time evolution of the ratio of the electron temperature
  $T_{yy}/T_{xx}$. Blue, red,  black and green curves are
  $\epsilon = 0.8, 0.4, 0.1, 0.0$, respectively.} 
  \label{fig:fig6}
 \end{center}
\end{figure}

\begin{figure*}
 \begin{center}
 \includegraphics[width=14cm]{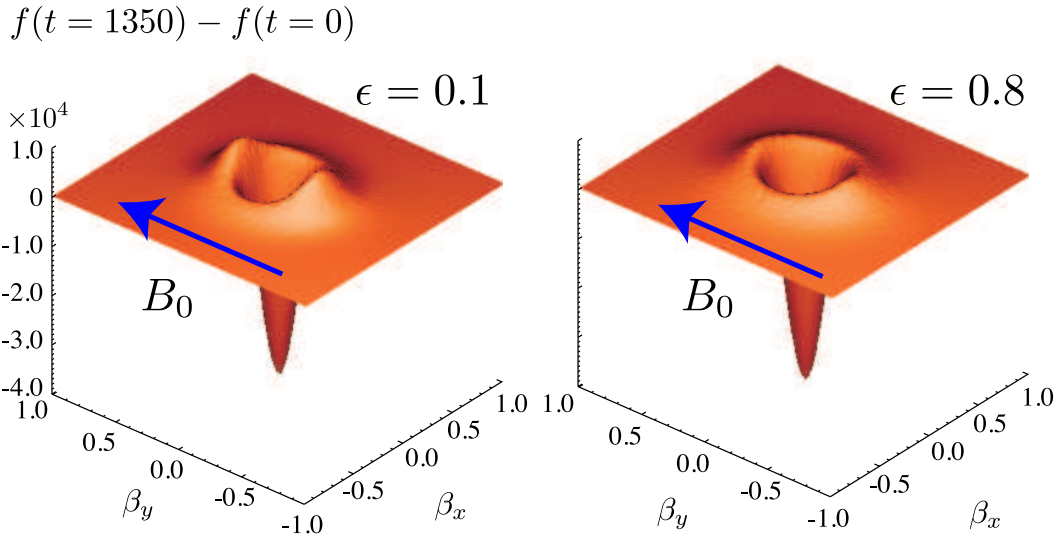}
  \caption{Deviation of the electron distribution function at
  $t\omega_{pe}=1350$ from the initial state at $t\omega_{pe}=0$ in
  $\beta_x-\beta_y$ plane. Blue arrows denote the direction of the uniform
  magnetic fields $\bmath B_0$.
  Left and right panels denote for $\epsilon = 0.1$ and $0.8$, respectively.} 
  \label{fig:fig7}
 \end{center}
\end{figure*}

To see the anisotropy more in detail, we perform the Fourier analysis
of the magnetic fields. Top panels of Figure~\ref{fig:fig3} show the
power spectral density (PSD) of the turbulent magnetic field
${|\widehat{\bmath B_t}/B_0|^2}$ at $t\omega_{pe}=1350$. 
It can be seen that the anisotropic turbulence preferentially transports the magnetic 
energy perpendicular to the uniform magnetic fields.
The PSD is larger for a larger $\epsilon$, suggesting that the larger
amplitudes of the turbulence lead to an efficient electron heating. 
The lower panels of Figure~\ref{fig:fig3} show  1-dimensional
cumulative power spectrum densities of the turbulent magnetic fields
at $t\omega_{pe}=1350$. The cumulative spectrum is defined as $|\widehat{
\bmath B_t}(k_x)|^2=\sum_{k_y} |\widehat{\bmath B_t}(k_x, k_y)|^2$
(thick solid curves) and $|\widehat{\bmath B_t}(k_y)|^2=\sum_{k_x}
|\widehat{\bmath B_t}(k_x, k_y)|^2$ (thin solid). 
The summation in $k$-space is performed in the rage of $0< k_x, k_y <
10$ to reduce thermal noises.
Dashed curves show the initial turbulent spectrum, which is
isotropic in $k_x$-$k_y$ space.
From these panels, the turbulent magnetic fields are shown to proceed
via the forward cascade perpendicular to the uniform magnetic fields. 
The PSD in the perpendicular direction has a clear power-law
distribution. Also the PSD in parallel direction also deviates from the
initial (gaussian) distribution for a larger $k$. It suggests that the
turbulent energy is mainly transported in perpendicular direction,
but a some part of the energy (a few percent) is transported in the
parallel direction.

To visualize the time evolution of the anisotropy, we show the time
evolution of a quantity $\theta$ in Figure~\ref{fig:fig4},
\begin{equation}
\tan^2\theta = \frac{\int \mathrm{d}k_x\mathrm{d}k_y~
 k_x^2 |\widehat{\bmath B_t}|^2}
 {\int \mathrm{d}k_x\mathrm{d}k_y~
 k_y^2 |\widehat{\bmath B_t}|^2},
\end{equation}
\citep[see][]{1983JPlPh..29..525S, 2008PhPl...15j2305S}.
 As already mentioned, the turbulent eddies interact each other with
 time more frequently for a larger $\epsilon$, transferring the
 turbulent energy to smaller scales. 
When $t \omega_{pe} \lesssim 100$, the
anisotropic cascades proceed faster for a larger $\epsilon$ because the
cascade rate is an increasing function of the fluctuation.
The turbulent energy is then transferred to the smaller
scale.  
After that, the anisotropy $\theta$
saturates almost at a constant level. 
The saturation level is shown to be highest for 
 $\epsilon  = 0.4$ among the computed models. This is because too much initial turbulent
 energy (like $\epsilon = 0.8$) disturbs the direction of the uniform magnetic fields,
 acting to smearing out the anisotropy.

\subsection{Electron Heating}
\begin{figure*}
 \begin{center}
 \includegraphics[width=13cm]{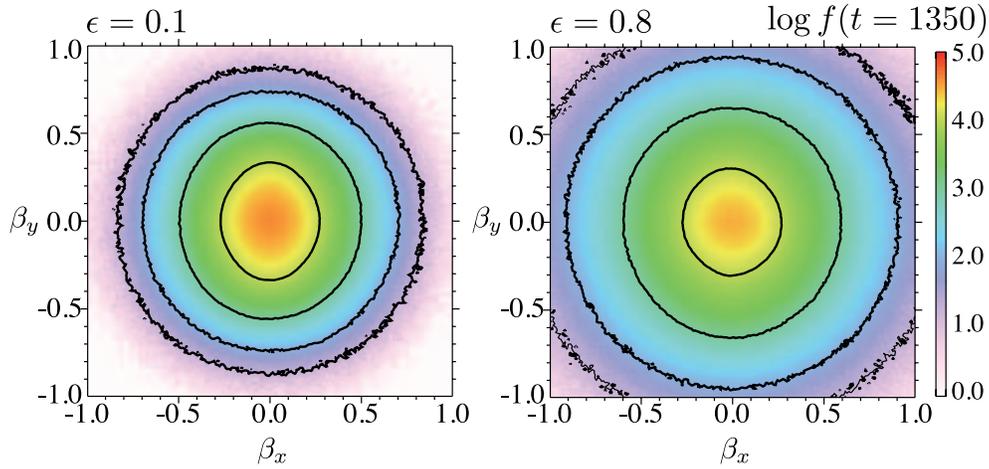}
  \caption{Electron distribution function at
  $t\omega_{pe}=1350$ for $\epsilon = 0.1 $ (left) and for $\epsilon =
  0.8$ (right) on the plane of 
  $\beta_x-\beta_y$.}
  \label{fig:fig8}
 \end{center}
\end{figure*}
Now we are in a position to 
  evaluate the temperature of electrons heated by the particle-wave interactions 
mentioned above. In doing so, we utilize the energy-momentum tensor as
\begin{equation}
 T^{\mu \nu} = \frac{\int \mathrm{d}^3\bmath p~ f(\bmath p) p^\mu
  p^\nu/e}{m_e\int \mathrm{d}^3 \bmath p~f(\bmath p)},\label{eq:Ttensor}
\end{equation}
where $e$, $p^\mu$ and $f$ are the electron energy, the electron four
momentum, and the electron distribution
function, respectively. 
 It should be noted that the temperature obtained from this equation
is evaluated in the observer frame, while it should be naturally defined in the
comoving frame. However this simplification is good enough in our case because 
 the averaged velocity of electrons in mostly random motions is much smaller 
than the speed of light.

Figure \ref{fig:fig5} shows the time evolution of the electron
temperature. 
It can be seen that the electron temperature parallel to the uniform
magnetic field ($T_{yy}$, solid curves) increases with time faster than
the perpendicular one ($T_{xx}$, dashed curves), and then $T_{yy}$ becomes larger than
$T_{xx}$. 
 It is also shown that the electrons are
 heated more rapidly for a larger $\epsilon$ because the turbulent energy cascade 
proceeds faster.
Figure~\ref{fig:fig6} shows the time evolution of the ratio of
$T_{yy}$ and $T_{xx}$. Before $t\omega_{pe}<200$,
the anisotropy of the electron temperature increases with time faster
for a larger $\epsilon$. In this epoch, 
the electron is preferentially heated in the direction 
 parallel to the uniform magnetic fields. Later on, the ratio 
 approaches to a nearly constant value. The saturation level is highest for 
$\epsilon = 0.4$ by the same reason as mentioned above (e.g., Figure \ref{fig:fig4}).

Figure~\ref{fig:fig7} shows the deviation of the electron distribution
function at $t\omega_{pe}=1350$ from the initial state at
$t\omega_{pe}=0$ for $\epsilon =
0.1$ (left panel) and $\epsilon = 0.8$ (right panel) on the plane of
$\beta_x-\beta_y$ with $\beta_i \equiv
v_i/c$ being the three velocity normalized by the speed of light. 
It can be seen that the distribution function is strongly distorted in 
the $y$-direction for $\epsilon = 0.1$. 
Considering the cascading process in perpendicular direction and the
electron heating along the uniform magnetic fields, a Landau resonance
($\omega=k_y v_y$) is the main mechanism for the electron
heating. The electromagnetic waves generated by the turbulent motion can
interact with electrons, which results in the Landau resonance.

\begin{figure}
 \begin{center}
 \includegraphics[width=8cm]{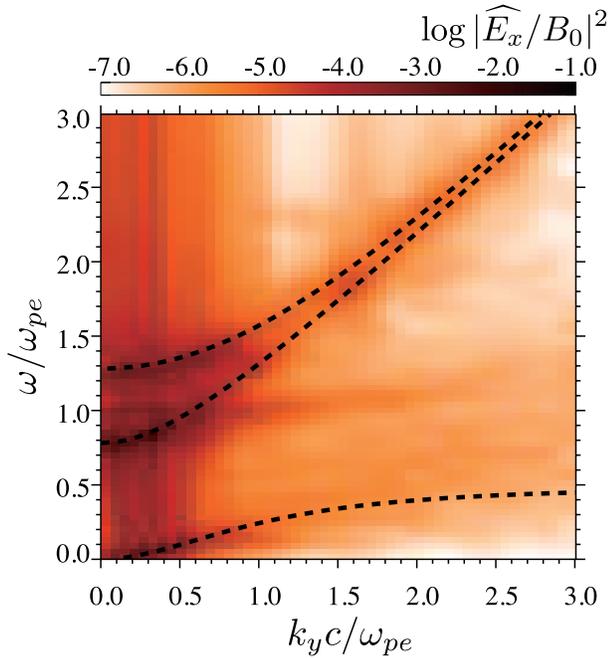}
  \caption{Dispersion relation of parallel propagating waves of
  $E_x$. The color shows the numerical result for $\epsilon =
  0.1$. The dashed curves represent the dispersion relations for the
  $R$-mode (top and bottom) and $L$-mode (middle) obtained from the linear analysis.}
  \label{fig:fig9}
 \end{center}
\end{figure}

 Figure \ref{fig:fig8} shows the electron distribution function at
$t\omega_{pe}=1350$ for $\epsilon = 0.1$ (left) and for $\epsilon = 0.8$
(right) on the plane of $\beta_x-\beta_y$. 
As mentioned above, the distribution function is distorted in parallel
direction because of the electron heating due to the Landau resonance.
 It should be noted that in both cases, the electrons
with higher 
energies (i.e., a larger $\beta$) are shown to get energies also in the direction 
 perpendicular to the uniform magnetic fields,
 thus making the distribution function almost isotropic (see also Figure \ref{fig:fig2}  
 that also shows the decrease in the parallel component of the magnetic field). 
This indicates that not only the Landau resonance, but also the
cyclotron resonance ($\omega = k_y v_y + \omega_{ce}$) plays a role of
the electron heating and the
dissipation of the turbulent magnetic energy \cite[e.g.,][]{2008PhPl...15j2305S}. 
While the direction of the electron heating is parallel to $B_0$ for
 the Landau resonance, there is an energy exchange between the waves and the
perpendicular motion of the particles when the cyclotron resonance
occurs \cite[see, e.g.,][]{1993tspm.book.....G}. 
It is also interesting to note that the energy dissipation rate of the
cyclotron resonance obtained in
this simulation is almost comparable to that of the Landau resonance, 
which has been known to be the case only in the linear regime \citep{2005spr.book.....M}.
 The dispersion relation of parallel propagating waves of $E_x$ is
shown in Figure \ref{fig:fig9} (color contour).
The dashed curves denote the dispersion relations for the $R$-mode (top
and bottom curves) and $L$-mode (middle curve) obtained from the linear
analysis. Although the $L$-mode waves are generated by the
turbulent motion, it is noted that they do not exchange
energies with electrons since their polarization is in an opposite
direction to the mean magnetic fields. The $L$-mode electromagnetic
waves can interact mainly with
ions but they are now assumed to be immovable as already mentioned. 
Thus the turbulence is
 maintained by the $R$-mode (whistler) waves in the electron scale.

Comparing to the results of the 
linear analysis (dashed lines), the dispersion relation
obtained from the numerical results is not so clear, mainly because
 the large amplitudes of the turbulent magnetic fields injected at
the initial state proceed forward cascades. By boldly
extrapolating the agreement with the linear analysis with respect to the
$R$-mode wave (whistler wave) for $k_y c/\omega_{pe} \lesssim 1.0$
(bottom line), 
the $R$-mode wave in the higher $k_y$ regime would be 
 damped due to the cyclotron resonance. The findings of this paper (i.e., 
the occurrence of the heating both in the parallel and 
perpendicular direction as well as the presence of the $R$-mode damping) 
suggest that the Landau resonance and the cyclotron resonance would be pivotal 
 factors for the dissipation of the magnetic energy and 
the resulting electron heating.

\section{Summary and Discussions} \label{sec:conclusion}
Bearing in mind the application to the outer crust of
the NS, we investigated the turbulent magnetic field decay 
by means of the fully relativistic 2-dimensional PIC
simulations. In the numerical simulations, the initial magnetic fields were
 set to be composed both of the uniform magnetic fields that model the global 
 fields penetrating the NS and of 
the turbulent magnetic fields that would be originated from the Hall cascade 
of the large-scale turbulence.
We showed that the turbulent whistler cascade transports 
the magnetic energy preferentially in the direction perpendicular to the uniform
magnetic fields.
 It was also found that the distribution function of electrons 
becomes anisotropic because electrons with lower energies are 
predominantly heated in the direction parallel to the uniform magnetic fields due to 
the Landau resonance, while electrons with higher energies are heated mainly 
by the cyclotron resonance that makes the distribution function isotropic
for the high energy tails. Furthermore we pointed out that the degree of anisotropy 
takes maximum as a function of the initial turbulent magnetic energy.
 This is because too much initial turbulent
 energy disturbs the direction of the uniform magnetic fields,
 acting to smearing out the anisotropy.
 The findings of this paper 
suggest that the particle-wave interactions via the 
Landau resonance and the cyclotron resonance, alternative to the conventional
 particle-particle
 collisions via the Ohmic dissipation, would be a pivotal dissipation mechanism
 in the outer crust of NSs.

The initial turbulent magnetic fields assumed in this study are relatively 
larger compared to the previous work \citep{2008PhPl...15j2305S,2008GeoRL..3502104G}. 
For $\epsilon < 0.4$ in our simulation, the obtained results of the anisotropic cascade 
and the electron heating are basically consistent with the previous studies.
 As already mentioned, the anisotropy of the temperature decreases as
$\epsilon$ increases furthermore since the turbulent fields disturb the direction of 
the uniform magnetic fields. Although the temperature becomes nearly isotropic, 
  it should be noted that the cascading process itself is still anisotropic. 
 According to \cite{Cho04,Cho09}, the small eddy (with size $\sim
l$) of the 
turbulent magnetic fields $\bmath  b_l$ interacts not with the global magnetic fields 
$B_0$, but with the 'local' mean magnetic fields $\bmath B_L$.
Thus even when the turbulent magnetic field energy is much smaller than that of the
uniform magnetic fields (especially $B_0 = 0$), the cascading process should be locally 
anisotropic along $\bmath B_L$. Such diffusion processes lead to the
anisotropic heating of electrons, however the resulting electron temperature is 
in average isotropic, which should be the case obtained for high $\epsilon$ 
in our simulation.

In the conventional model of the magnetic field decay, the
magnetic fields are considered to be dissipated through the collisional process, 
namely via the Ohmic dissipation. While in this paper, 
we proposed that the magnetic fields are 
dissipated due to the Landau and cyclotron resonances, leading to the plasma heating 
in the collisionless regime. 
Since the wave-particle interactions occur in a very small scale
($\sim$ gyro radius), the corresponding timescale ($\sim
\omega_{ce}^{-1}$) is much shorter than that of the Ohmic dissipation,
 and it is instantaneous compared to the expected magnetic field decay timescale 
 in the NSs. 
This means that the decay timescale should be 
 determined by the cascading processes of the magnetic fields {\it above}
 the electron inertia scale. Although the PIC simulations have an advantage in its 
 capability to determine the dissipation processes consistently, 
it is still computational too expensive to perform the PIC simulations covering
 over such a wide spacial range required to estimate the timescale. For the purpose, 
 we think it important to perform the EMHD simulations including a phenomenological
  diffusivity which is adjusted to mimic the dissipation obtained in this study. 
Although this is apparently beyond the scope of this paper, we regard it as one of 
the most important tasks, which we plan to investigate as a sequel of this study.

So far there have been extensive work
 that focuses on the origin of the observed anisotropy in the NSs' surface temperature
 \cite[e.g.,][]{geppert1,pons07,agui,pons1}.  
 Unfortunately it is hard for us to do so immediately because  
the anisotropy obtained in the current simulation 
is confined in a very small scale (collisionless scale). To clarify how 
the particle-particle collisions in the more larger scales redistribute the electron 
distribution function and the resulting global temperature could be, one may need a new 
simulation technique which bridges the PIC simulation and the global (E)MHD simulations, albeit not an easy job at present.
 In addition to the NSs' surface, we speculate that the anisotropy due to the Landau 
resonance could be important also in 
 the pulsar atmosphere because the plasma density is so small that the plasma there is
 expected to be globally collisionless.
The consideration of the electron heating due to the Landau and
cyclotron resonances might make the origin of the observed temperature anisotropy 
of NSs \citep{2007astro.ph..2426Z, 2007Ap&SS.308..181H, 2009PASJ...61S.387N} 
less mysterious, which we are going to study one by one as an extension of this study.

\acknowledgments

Numerical computations were carried out on Cray XT4 at Center for
Computational Astrophysics, CfCA, of National Astronomical Observatory
of Japan and on Fujitsu FX-1 at JAXA Supercomputer System (JSS) of Japan
Aerospace Exploration Agency (JAXA).
This study was supported in part by the Grants-in-Aid for the Scientific Research 
from the Ministry of Education, Science and Culture of Japan
(Nos. 19540309 and 20740150).

\end{document}